\documentclass[twoside]{article}
\usepackage{fleqn,espcrc2,epsfig}

\newcommand{\AmS}{{\protect\the\textfont2
  A\kern-.1667em\lower.5ex\hbox{M}\kern-.125emS}}

\title{New anomalous exchange in Regge phenomenology and hard diffraction}

\author{
\underline{Nikolai I. Kochelev}%
\address[JINR]{BLTP, JINR, 141980 Dubna,  Russia }%
	\thanks{e-mail: kochelev@thsun1.jinr.ru},
Dong-Pil Min%
\address[SNU]{Department of Physics,
     Seoul National University, Seoul 151-742, Korea}%
	\thanks{e-mail: dpmin@mulli.snu.ac.kr},
Yongseok Oh%
\address[YU]{Institute of Physics and Applied Physics,
     Yonsei University, Seoul 120-749, Korea}%
	\thanks{e-mail: yoh@phya.yonsei.ac.kr},
Vicente Vento%
\address[UV]{Departament de F\'{\i}sica Te\`orica and Institut de
     F\'{\i}sica Corpuscular, Universitat de Val\`encia-CSIC E-46100,
     Burjassot (Valencia), Spain}%
	\thanks{e-mail: Vicente.Vento@uv.es},
Andrey V. Vinnikov%
\address[FESU]{Far Eastern State University, Sukhanova 8, GSP, Vladivostok,
     690660 Russia}%
	\thanks{e-mail: vinnikov@thsun1.jinr.ru}}

\begin{document}

\begin{abstract}
A new mechanism for hard diffraction based on the anomalous $f_1$
trajectory exchange, which we identify as the odd signature partner
of the Pomeron, is suggested.
We calculate the contribution of the $f_1$ exchange to elastic and
dissociative electromagnetic production of vector mesons and
show that it gives a dominant contribution to the differential cross
sections at large momentum transfers.
 \vspace{1pc}
\end{abstract}

\maketitle

\section{Introduction}

Understanding the mechanisms of diffraction at large energies is one of
the main topics in QCD \cite{land}.
This interest comes from the possibility that the fundamental problems
of QCD in the diffractive processes are related to the perturbative and
nonperturbative quark-gluon interactions.
One of such problems is the role of the complex structure of QCD vacuum
in parton scattering.
A well-known fact is that at large energies the dominant contribution to
the parton-parton scattering cross sections comes from the exchange which
has the vacuum quantum numbers.
At small momentum transfers this exchange can be successfully described
by the famous Pomeron which has $P=C=+1$ and even signature.
If the virtualities of the quarks and gluons are large enough, then the
hard BFKL Pomeron, based on pQCD, would play an important role.
When the virtualities of the partons are small, the QCD vacuum structure
becomes very important, where the soft Pomeron contribution is dominant
phenomenologically.
The relations between the soft Pomeron properties and the QCD vacuum
structure are widely under discussion \cite{KL1}.
The most interesting idea is relating the unusual properties of the
Pomeron trajectory, e.g., its high intercept and small slope, with the
existence of the {\it scale} anomaly in QCD \cite{KL1}.

There is another class of diffractive process, i.e., the hard diffraction
at large $t$ and small virtualities of initial and final particles, which
has been much less studied experimentally and theoretically.
In this process, colourless exchanges between partons are involved
having large momentum transfers.
The simplest example of such reactions is elastic hadron-hadron scattering
at large $t$.
It should be mentioned that the first data from ISR on $pp$ elastic
scattering at large $t$ have already shown some unexpected results,
e.g., only one diffractive minimum was observed instead of the set of
minima anticipated from the multi-Pomeron exchange picture.
In order to explain such phenomena, the Odderon exchange hypothesis,
which has $P=C=-1$ and intercept $\alpha_O(0) \approx 1$,
has been suggested \cite{DL79}.
The peculiarity of such exchange is in its negative charge parity, which
should lead in particular to the difference between the $pp$ and the $p\bar p$
scattering cross sections at large energies.
However it was also claimed that the modern data on $pp$ and $p\bar p$
elastic scattering could be explained without introducing the Odderon
exchange at least at $t=0$ \cite{white}.

New data on large $t$ hard diffraction are now available from HERA
\cite{ZEUS1,ZEUS2,H1}.
These data show unusual energy dependence of the cross sections for
vector meson electromagnetic production and thus call for new ideas as
they are difficult to explain within the conventional mechanisms
based on the soft Pomeron exchange.
Recently we suggested a mechanism \cite{KMOV00} to understand these data
by introducing a new anomalous $f_1$ exchange in Regge theory,
which corresponds to an effective trajectory
with a high intercept $\alpha_{f_1}^{}(0) \approx 1$ and small slope
$\alpha_{f_1}^\prime \approx 0$.
This exchange has the quantum numbers of the flavor-singlet axial-vector
$f_1(1285)$ meson, i.e., $P = C = +1$ and odd signature, and can be
considered as the natural partner of another anomalous Regge trajectory,
i.e., the Pomeron, which has even signature and the same parities
$P = C = +1$.
Moreover we have shown that the $f_1$ exchange allows us to understand
the anomalous behavior of the spin-dependent structure function
$g_1(x)$ at low $x$ and the universal behavior of the differential cross
sections for elastic $pp$ and $p\bar p$ scattering at large momentum
transfers.
The quantum numbers of this trajectory ($P=C=+1$) are different from
those of the Odderon ($P=C=-1$) \cite{DL79} and can be verified by
experimental measurements on several spin-dependent
cross sections \cite{KMOV00,OKMV00}.

\section{Elastic electromagnetic production of vector mesons and
\boldmath{$f_1$} exchange}

Among the possible effective quark-quark interactions, there are only
two kind of vertices, $\gamma_\mu \otimes \gamma_\mu$ and
$\gamma_\mu \gamma_5 \otimes \gamma_\mu \gamma_5$, which survive at
high energies and whose Lorentz structure does {\it not\/} lead to the
quark helicity-flip process at large energy $s$ and small momentum
transfer $-t \ll s$. The spin-flip property of the quark-quark scattering 
is very important
for the high energy behavior of the corresponding scattering amplitude
because any spin-flip amplitude should be suppressed by additional
factors $t/s$ due to total angular momentum conservation.
Therefore only the above mentioned Lorentz structures can lead to
non-vanishing cross sections at large energies in the forward scattering
limit.

The vector-like structure $\gamma_\mu \otimes \gamma_\mu $ has been
suggested many years ago as a possible effective interaction induced by
the soft Pomeron exchange \cite{LN87}.
This structure has enjoyed much success in the description of many
hadron-hadron, lepton-hadron and photon-hadron reactions at small
$|t|$ region ($|t| \leq 1$ GeV$^2$), as well as their total cross sections,
once the phenomenological Pomeron trajectory and form factors are used.
However the possible existence of the other important structure
$\gamma_\mu \gamma_5 \otimes \gamma_\mu \gamma_5$ has not been noticed
seriously until Ref. \cite{KMOV00}.
It should be emphasized that the effective interactions for both
anomalous exchanges have the structure of the product of two vector or
axial-vector currents: $J_\mu^P \otimes J_\mu^P$ for the Pomeron and
$J_{5\mu}^a \otimes J_{5\mu}^a$ for the $f_1$ exchange.
The similarity of the Lorentz structure of both exchanges has probably a
deep origin.
The Pomeron current is expected to be connected with the non-conserved
dilaton current whose divergence is given by the scale anomaly,
and the $f_1$ anomalous exchange is related to the non-conserved axial
vector current whose divergence is given by the axial anomaly \cite{KMOV00}.
The connection between these anomalous exchanges would also be natural
in SUSY QCD where only some definite chiral combinations $G^2\pm i G
\widetilde{G}$ correspond to superfields \cite{KZ90}, which leads to an
expectation that these trajectories are degenerate in SUSY QCD.

On the hadronic level the anomaly manifests itself in unusual
decay rates.
The famous example is the axial anomaly  in $\pi^0\to \gamma\gamma$
decay due to triangle graph contribution to the transition amplitude.
The large mass of the $\eta^\prime$ meson is also related to the large
mixing of this flavor singlet with gluons due to axial anomaly.
Recently it was suggested that abnormal properties of Pomeron trajectory
would be explained by the large mixing between glueball and $q\bar q$
trajectories \cite{kaidalov}.
In the axial vector channel the lowest meson state with the appropriate
quantum numbers, $I=0$ and $P=C=+1$, is the $f_1(1285)$.
This flavor-singlet axial vector meson can couple to the quarks
through mixing with a two gluon state due to triangle anomaly.
In accordance with the Landau-Yang theorem the coupling of $1^{++}$
state to two gluon state vanishes for on-shell gluons.
Therefore the intermediate gluons should be highly virtual to give
a contribution to the $f_1qq$ coupling and we thus consider
$f_1qq$ vertex as {\it point-like}.

The diffractive electromagnetic production of vector mesons is
one of the appropriate processes for testing different perturbative
QCD and QCD-inspired nonperturbative models for quark-quark scattering.
Recently new experimental data on photo- and electro-production of
$\rho$, $\omega$ and $\phi$ mesons at large energyies $W \approx 100$
GeV, and large momentum transfers, up to $-t \approx 2$ GeV$^2$ in elastic
production and up to $-t \approx 11$ GeV$^2$ in low-mass dissociation,
have been reported \cite{ZEUS1,ZEUS2}.
At small $|t|$ and large $W$, the fixed-target photoproduction data have
been described successfully within the soft Pomeron exchange models with
the universal Pomeron trajectory $\alpha_P^{\rm soft}(t)=1.08+0.25t$.
But the reported new data violate this universality at large $|t|$ region
\cite{ZEUS1}.
In order to describe such observations in terms of Regge trajectories, an
additional Pomeron, called {\it hard\/} Pomeron, with the trajectory
$\alpha_P^{\rm hard} = 1.4 + 0.1t$ has been proposed \cite{DL00}.
However it is not yet certain whether this approach can also describe the
energy independence of the elastic $pp$ and $p\bar p$ cross sections at
large energies and momentum transfers.

In a recent paper \cite{KMOV00} we have shown that the anomalous $f_1$
exchange can describe both the elastic hadron-hadron and $\rho/\phi$
photoproduction cross sections in a consistent way.
In the calculation of the $f_1$ exchange contribution to $\rho$ and
$\phi$ photoproduction, we have assumed a specific phenomenological form
factor in the $f_1^{} V \gamma$ vertex, which we would like now to relax.
Furthermore the strength of this vertex was fixed from the decay width
of $f_1 \to \gamma V$, which is a questionable procedure for
parameters to be used in high energy scattering where the quark
structure of the hadrons is important.
Moreover the absence of experimental data on the $f_1^{} \to \gamma
\omega$ decay did not allow us to estimate the $f_1^{}$ exchange
contribution to $\omega$ meson photoproduction.

\begin{figure}[t]
\centering
\epsfig{file=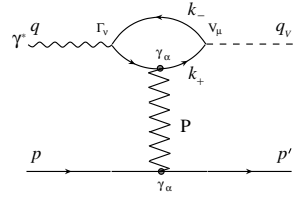,width=4cm}
\vskip -0.5cm
\caption{Pomeron exchange contribution to vector meson production.}
\label{fig:pomdiag}
\end{figure}

In this work we use a method similar to that used to calculate the Pomeron
exchange contribution to vector meson production in Ref. \cite{LN87}.
We then generalize the approach to include the $f_1$ exchange process
in the same production amplitude.
One important feature of this approach is that we have less parameters
for photoproduction and that it can be generalized straightforwardly
to electroproduction.
The Pomeron exchange contribution in the electromagnetic production of
vector mesons is presented in Fig. \ref{fig:pomdiag}.
The soft Pomeron contribution to transverse virtual photoproduction
cross section at large energy reads
\begin{eqnarray}
\frac{d\sigma^T_P}{dt}& = &\frac{27 m_V^3 \beta_u^2\beta_f^2
\Gamma_{e^+e^-}^V}
{\pi \alpha_{\rm em}}
\left( \frac{s}{s_0} \right)^{2 \alpha(t) - 2}
\nonumber\\ & & \mbox{} \times
\left[ \frac{F_1(t)\sqrt{1-t/2m_V^2}}{(Q^2 + m_V^2 - t)}\right]^2
F(t,Q^2),
\end{eqnarray}
where the factor $F(t,Q^2)$ accounts for the nonlocality of the
Pomeron-quark vertex.

\begin{figure}[t]
\centering
\epsfig{file=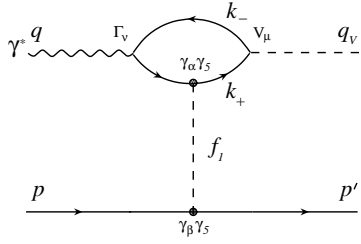, width=5cm}
\vskip -0.5cm
\caption{$f_1$ exchange contribution to vector meson production.}
\label{fig:f1diag}
\end{figure}

\begin{figure}[t]
\centering
\epsfig{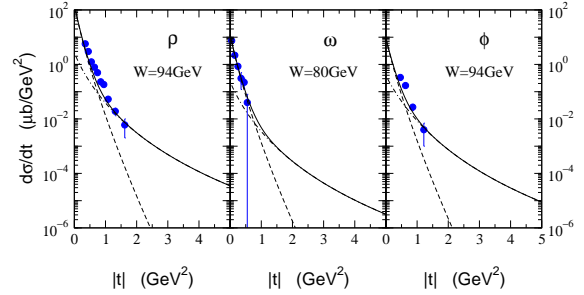}
\vskip -0.5cm
\caption{The differential cross sections for (a) $\rho$ meson
at $W=94$ GeV, (b) $\omega$ meson at $W=80$ GeV and (c) $\phi$ meson
photoproduction at $W = 94$ GeV.
The dashed and dot-dashed lines are the Pomeron and the $f_1$ exchange
contributions, respectively, while the solid lines show the results by
including the two exchanges.
The experimental data are from Refs. \cite{ZEUS1,ZEUS96}.}
\label{fig:diff0}
\end{figure}

The contribution of the $f_1$ exchange to vector meson electromagnetic
production (Fig. \ref{fig:f1diag}) is obtained in a similar manner as
in the calculation of the Pomeron exchange amplitude.
The main difference lies on the substitution of $\gamma_\alpha$ matrix
in the effective Pomeron-quark vertex by $\gamma_\alpha \gamma_5$ in
the effective $f_1$-quark vertex.
Note that we do not have any new free parameters for
this exchange since the coupling constant $g_{f_1^{}ff}$ is fixed by
\begin{equation}
g_{f_1ff}^{} = g_{f_1NN}^{},
\label{gnn}
\end{equation}
which can be obtained using a constituent quark picture for the nucleon wave
function.
It should be mentioned that this relation is different from the additivity
of the Pomeron coupling, i.e., $g_{PNN}^{} \approx 3g_{Pqq}^{}$.
What Eq. (\ref{gnn}) implies is that the axial vector coupling is
sensitive only to the spin of the particle.
In our calculation we use $g_{f_1NN}^{} = 2.5$ that is obtained from the
analyses of the $f_1$ contribution to the proton spin
(see Ref. \cite{KMOV00}).
Then the final form for the $f_1$ exchange contribution to transverse
virtual photoproduction cross section is (for $\alpha_{f_1}(0)=1 $ and
$\alpha_{f_1}^\prime=0$)
\begin{eqnarray}
\frac{d\sigma^T_A}{dt}& =& \frac{3 m_V^3 g_{f_1^{}NN}^4
\Gamma_{e^+e^-}^V}{\pi \alpha_{\rm em}}
\nonumber \\ && \mbox{} \times
\left[ \frac{F_{f_1^{}NN}(t)\sqrt{1-t/2m_V^2}}{(Q^2 + m_V^2 -
t)(t-m_{f_1^{}}^2)}\right]^2,
\label{cross}
\end{eqnarray}
where $F_{f_1 NN}$ is the flavor singlet axial vector form factor of the
nucleon.
In Fig. \ref{fig:diff0} we present the calculated differential cross
sections for vector meson ($\rho$, $\omega$, $\phi$) photoproduction
together with the experimental data \cite{ZEUS1,ZEUS96}.
The parameters are $\beta_u=\beta_d=2$ GeV$^{-1}$, $\beta_s=1.5$
GeV$^{-1}$ and $s_0=4$ GeV$^2$.

One can see that our model reproduces the experimental data 
successfully.
It should be mentioned that the $f_1$ contribution dominates over that
of the Pomeron in the large $|t|$ region.
This dominance is closely related to the slow decrease of the
flavor singlet axial vector form factor of the nucleon
with respect to $|t|$ as compared with that of the electromagnetic
form factor, and to the different slopes of the trajectories.

\section{\boldmath{$f_1$} contribution to vector meson photoproduction
in proton dissociation}

\begin{figure}[t]
\centering
\epsfig{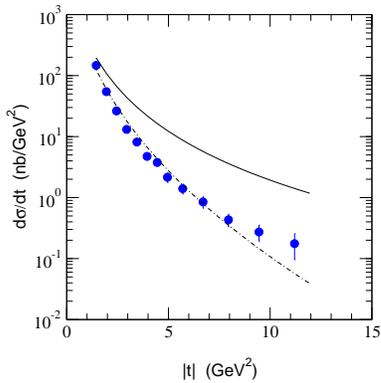}
\vskip -0.5cm
\caption{The differential cross section for $\rho$ meson photoproduction
in proton dissociation process at $W = 100$ GeV.
The solid and dashed lines are the $f_1$ exchange contribution with zero
slope and with $\alpha^\prime_{f_1}=0.025$ GeV$^{-2}$, respectively.
The experimental data are from Ref. \cite{ZEUS2}.}
\label{fig:ZEUSdis}
\end{figure}

The recent report on vector meson photoproduction at large $t$ in
proton dissociation by the ZEUS Collaboration \cite{ZEUS2} shows a
surprising behaviour of the differential cross sections.
The differential cross section data have very hard $t$ dependence.
The observed data have $1/t^3$ dependence instead of the $1/t^4$
dependence which is expected from the pQCD approach.
Within our model the dominant contribution to the proton dissociation
at large $t$ comes from the $f_1$ exchange between the proton quarks and
the quark-antiquark pair arising from the photon.
It was mentioned that it would be a good approximation to consider the
$f_1$-quark vertex as a point-like vertex.
Thus the main difference between the $f_1$ contributions to elastic
[Eq. (\ref{cross})] and dissociative production of vector mesons
is the absence of the proton form factor in the formula for the
differential cross section,
\begin{eqnarray}
\frac{d\sigma^{diss}_{f_1}}{dtdx}&=&\sum_iq_i(x,t)\frac{3 m_V^3 g_{f_1^{}NN}^4
\Gamma_{e^+e^-}^V}{\pi \alpha_{\rm em}}
\nonumber\\ && \mbox{} \times
\left[ \frac{\sqrt{1-t/2m_V^2}}{(Q^2 + m_V^2 -
t)(t-m_{f_1^{}}^2)}\right]^2,
\end{eqnarray}
where $q_i(x,t)$ is the distribution of $i$-quark ($i=u,d,s$) in the
proton, $x=-t/(M_x^2-t)$ and $M_x$ is the mass of the dissociative
system.
In Fig. \ref{fig:ZEUSdis} we show the results of our calculation
together with preliminary ZEUS data \cite{ZEUS2}.
We have used the ZEUS cut $x>0.01$ and LO GRV98 \cite{GRV98}
partonic distributions.
The agreement with the experimental data is rather good.
The deviation at large $t$ may be due to the manifestation of non-zero
(although small) slope of the $f_1$ trajectory (dashed line) or the
deviation from the locality of the $f_1$-quark vertex.

\section{Summary}

We have calculated the $f_1$ anomalous trajectory contribution to
vector meson electromagnetic production and have shown its important
role in differential cross sections at large momentum transfers.
The proton dissociation at large $t$ is one of the processes
where the contribution from the $f_1$ exchange is very large.
We have shown that the additional anomalous trajectory can explain
preliminary ZEUS data reasonably.

N.I.K. would like to express his sincere gratitude to the
Organizing Committee of the Diffraction 2000 Conference and especially
to R. Fiore for the kind invitation and financial support.
He is also grateful to many participants of the conference for fruitful
discussions.
This work was supported in part by the Brain Korea 21 project of Korean
Ministry of Education.

\end{document}